\documentclass[12pt]{article}

\usepackage[spanish,english]{babel}
\usepackage{amsfonts}
\usepackage{amssymb}
\usepackage{graphicx}

\newcommand{\be}{\begin{equation}}
\newcommand{\ee}{\end{equation}}
\newcommand{\bea}{\begin{eqnarray}}
\newcommand{\eea}{\end{eqnarray}}

\usepackage{a4wide}
\begin{document}

\begin{titlepage}


\bigskip

\begin{center}

{\bf{\LARGE Particles and energy fluxes\\ from a CFT
 perspective}}

\bigskip
\bigskip\bigskip
 A.  Fabbri$^{a,c}$, \footnote{e-mail: \sc
fabbria@bo.infn.it} J. Navarro-Salas$^b$ \footnote{\sc
jnavarro@ific.uv.es} and
 G.J. Olmo$^{b}$ \footnote{\sc gonzalo.olmo@ific.uv.es}

\end{center}

\bigskip%

\footnotesize \noindent
 a) Dipartimento di Fisica dell'Universit\`a di Bologna and INFN
 sezione di Bologna, Via Irnerio 46,  40126 Bologna, Italy
 \newline
 b) Departamento de F\'{\i}sica Te\'orica and
    IFIC, Centro Mixto Universidad de Valencia-CSIC.
    Facultad de F\'{\i}sica, Universidad de Valencia,
        Burjassot-46100, Valencia, Spain
\newline
c) GReCO, Institut d'Astrophysique de Paris, CNRS, 98bis Boul. Arago,
75014 Paris, France

\bigskip

\bigskip

\begin{center}
{\bf Abstract}
\end{center}

We analyze the creation of particles in two dimensions under the
action of conformal transformations. We focus our attention on
Mobius transformations and compare the usual approach, based on
the Bogolubov coefficients, with an alternative but equivalent
viewpoint based on correlation functions. In the latter approach
the absence of particle production under full Mobius
transformations is manifest. Moreover, we give examples, using the
moving-mirror analogy, to illustrate the close relation between
the production of quanta and energy.



\end{titlepage}

\newpage

\section{Introduction}
One of the basic ingredients of quantum field theory in curved
spacetime \cite{birrelldavies} are the Bogolubov transformations.
These reflect the absence, in general, of a privileged vacuum
state, in parallel to the absence of global inertial frames. This
framework is general and can be applied to a large number of
physical situations, including flat spacetime backgrounds (like
the Unruh-Fulling effect \cite{birrelldavies}). On the other hand,
of particular physical interest are those field theories
possessing the spacetime conformal symmetry $SO(d,2)$, where $d$
is the dimension of the Lorentzian spacetime. This symmetry is
especially powerful in two dimensions, where the group $SO(2,2)$
can be enlarged to an infinite-dimensional group \cite{ginsparg}.
However, this $SO(2,2)$ subgroup, which includes dilatations,
Poincar\'e and special conformal transformations, still plays an
important role because it leaves the vacuum invariant
\cite{ginsparg}. From the point of view of Bogolubov
transformations this should imply that the $\beta$ coefficients
associated to them vanish. This is obvious for Poincar\'e and
dilatations: they do not produce any mixing of positive and
negative frequencies. However, this result is far from being
obvious for special conformal transformations.

In addition, a restriction to one of the two branches of special
conformal transformation  produces, in the context of moving
mirrors, a non-vanishing result \cite{daviesfulling,
birrelldavies}. Since Mobius transformations never produce local
energy fluxes this has been interpreted as a manifestation of the
fact that the production of quanta does not require presence of
energy \cite{daviesfulling, birrelldavies}. This claim has been
criticized in \cite{grove} using an explicit particle detector.

 The purpose of this note is
to clarify all the above issues. To this end we shall analyze the
phenomena of quantal production in a different way, more close to
the philosophy of Conformal Field Theory. In the new perspective,
the absence of particle production for the full set of Mobius
transformations (including the special conformal transformation)
is obvious from the very beginning, in sharp contrast to the usual
approach based on the Bogolubov transformations. We shall analyze
the corresponding moving-mirror analogy for special conformal
transformations (with one and two hyperbolic branches) to
illustrate, in an easy way, how the production of energy and
quanta are indeed closely related.

\section{Particle production and Bogolubov coefficients}
Let us first briefly review the definition of the Bogolubov
coefficients for the two-dimensional massless scalar field $f$
satisfying the wave equation  \be \nabla^2 f = 0 \ . \ee In
conformal gauge $ds^2=-e^{2\rho}dx^+dx^-$ we can decompose the
field into positive and negative frequencies using a mode
expansion :
 \be f= \sum_i \left( {\stackrel{\rightarrow}{a}}_i u_i(x^-) + {\stackrel{\rightarrow}{a}}^{\dagger}_i u_i^*(x^-)
 +  {\stackrel{\leftarrow}{a}}_i v_i(x^+) + {\stackrel{\leftarrow}{a}}^{\dagger}_i v_i^*(x^+)\right). \ee
  These modes must form an orthonormal basis under the
scalar product \be
\label{eq:scalarproduct}(f_{1},f_{2})=-i\int_{\Sigma}
d\Sigma^{\mu}\left (f_{1}\partial_{\mu}f_{2}^* -
\partial_{\mu}f_{1} f_{2}^* \right )\ , \ee where $\Sigma$ is an
appropiate Cauchy hypersurface.
 One can construct the Fock space from the commutation
relations \be
[{\stackrel{\rightarrow}{a}}_i,{\stackrel{\rightarrow}{a}}_j^{\dagger}]=\delta_{ij}
\ , \ee \be
 [{\stackrel{\leftarrow}{a}}_i,{\stackrel{\leftarrow}{a}}_j^{\dagger}]=\delta_{ij} \ . \ee The vacuum state $|0_{x} \rangle$ is defined
by \be {\stackrel{\rightarrow}{a}}_i|0_{x} \rangle = 0, \ \ \
{\stackrel{\leftarrow}{a}}_i|0_{x} \rangle = 0 \ ,\ee and the
excited states can be obtained by the application of creation
operators ${\stackrel{\rightarrow}{a}}_i^{\dagger},
{\stackrel{\leftarrow}{a}}_i^{\dagger}$ out of the vacuum. We can
perform an arbitrary conformal transformation
 \be
\label{cs}x^{\pm} \rightarrow y^{\pm}=y^{\pm}(x^{\pm}) \ , \ee and consider the
expansion
 \be f= \sum_j \left( {\stackrel{\rightarrow}{b}}_j \tilde{u}_j(y^-)
 + {\stackrel{\rightarrow}{b}}^{\dagger}_j \tilde{u}^*_j(y^-)+
{\stackrel{\leftarrow}{b}}_j \tilde{v}_j(y^+)
 + {\stackrel{\leftarrow}{b}}^{\dagger}_j
 \tilde{v}^*_j(y^+)\right) \ . \ee
As both sets of modes are complete, the new modes
$\tilde{u}_j(y^-)$, $\tilde{v}_j(y^+)$  can be expanded in terms
of the old ones:
    \begin{eqnarray}\label{Bog-modes}
    \begin{array}{c}
    \tilde{u}_j(y^-)=\sum_i \left( \alpha_{ji}u_i(x^-)
    +\beta_{ji}u_i^*(x^-)\right) \\
    \tilde{v}_j(y^+)=\sum_i \left( \gamma_{ji}v_i(x^+)
    +\eta_{ji}v_i^*(x^+)\right)
    \end{array}
      \end{eqnarray}
where  $\alpha_{ji}$, $\beta_{ji}$ , $\gamma_{ji}$ and $\eta_{ji}$
are called Bogolubov coefficients. These coefficients can be
evaluated by the following scalar products
    \begin{eqnarray}
    \begin{array}{cc}
    \alpha_{ji} = (\tilde{u}_j,u_i) & \beta_{ji} = -
    (\tilde{u}_j,u^*_i) \\
    \gamma_{ji} = (\tilde{v}_j,v_i) & \eta_{ji} = -
    (\tilde{v}_j,v^*_i)
    \ . \end{array}
    \end{eqnarray}

    The relation between creation and annihilation operators in the two basis is
    \begin{eqnarray}\label{a-in}
    \begin{array}{c}
    {\stackrel{\rightarrow}{b}}_j=\sum_i\left(
    \alpha_{ji}^*{\stackrel{\rightarrow}{a}}_i
    -\beta_{ji}^*{\stackrel{\rightarrow}{a}}^{\dagger}_i\right) \\
        {\stackrel{\leftarrow}{b}}_j=\sum_i\left(
    \gamma_{ji}^*{\stackrel{\leftarrow}{a}}_i
    -\eta_{ji}^*{\stackrel{\leftarrow}{a}}^{\dagger}_i\right)
    \end{array}
     \end{eqnarray}
    along with the corresponding ones for
    ${\stackrel{\rightarrow}{b}}_j^{\dagger}$ and ${\stackrel{\leftarrow}{b}}_j^{\dagger}$.
    Therefore the expectation value of the (right-mover sector) particle number
    operator ${\stackrel{\rightarrow}{N}}_j \equiv
     {\stackrel{\rightarrow}{b}}_j^{\dagger}{\stackrel{\rightarrow}{b}}_j$ is given by the
    expression
    \be \label{pnbog}
   \langle 0_{x}| {\stackrel{\rightarrow}{N}}_j|0_{x} \rangle =
   \sum_i |\beta_{ji}|^2 \ . \ee

If we consider Mobius transformations the quatities $\langle
0_x|N_{i}|0_x \rangle $ should vanish since, in a conformal field
theory, the vacuum is invariant under these transformations. This
means that the corresponding $\beta$ coefficients should also
vanish. An explicit analysis show that this is far from being
obvious (see later), despite of the fact that the invariance of
the vacuum under Mobius transformations is almost an axiom in CFT
\cite{ginsparg}. In next section we shall analyze the particle
production from a different perspective. We shall give a different
expression for $\langle 0_x|N_{i}|0_x \rangle $, in terms of which
the invariance of the vacuum under full Mobius transformation, one
of the basic cornerstones of CFT, will be manifest.

\section{Particle production and correlations}

Let us now show how to obtain an expression for $ \langle 0_{x}|
{\stackrel{\rightarrow}{N}}_j|0_{x} \rangle $ without introducing
explicitly the Bogolubov coefficients. Our starting point is the
two-point correlation function for the derivatives of the field
$f$ \be \label{correlatordfdf}\langle 0_x
|\partial_{\pm}f(x^{\pm})\partial_{\pm}f(x'^{\pm})|0_x \rangle =
-\frac{1}{4\pi}\frac{1}{(x^{\pm}-x'^{\pm})^2} \ . \ee Under
conformal transformations $x^{\pm} \rightarrow
y^{\pm}=y^{\pm}(x^{\pm})$, the above correlation functions
transform according to the rule for primary fields
\cite{ginsparg}: \be \label{eq:notation}\langle 0_x
|\partial_{\pm} f(y^{\pm})
\partial_{\pm} f(y'^{\pm}) |0_x \rangle
 =-\frac{1}{4\pi} \left(\frac{dx^{\pm}(y^{\pm})}{dy^{\pm}}\right)
 \left(\frac{dx^{\pm}(y'^{\pm})}{dy^{\pm}}\right)\frac{1}{(x^{\pm}(y^{\pm})-x'^{\pm}(y^{\pm}))^2}.
\ee These relations are fundamental to construct the normal
ordered stress tensor \\ \noindent $:T_{\pm\pm}:$, but also for the particle
number operator.
 In the coordinates $\{x^{\pm}\}$ 
  the normal ordered stress tensor operator
can be defined via point-splitting \be :T_{\pm\pm}(x^{\pm}):
=\lim_{x^{\pm}\to x'^{\pm }} :\partial_{\pm}f(
x^{\pm})\partial_{\pm}f(x'^{\pm }): \ , \ee where \be
:\partial_{\pm}f( x^{\pm})\partial_{\pm}f(x'^{\pm
}):=\partial_{\pm}f( x^{\pm})\partial_{\pm}f(x'^{\pm })+
\frac{1}{4\pi}\frac{1}{(x^{\pm}-x'^{\pm})^2}
 . \ee
Similar relations hold in the coordinates $\{y^{\pm}\}$. It is
easy to relate $:T_{\pm\pm}(y^{\pm}):$ with
$:T_{\pm\pm}(x^{\pm}):$ since \bea \label{normalorder}
:\partial_{\pm} f(y^{\pm})
\partial_{\pm} f(y'^{\pm}):
 &=& \frac{dx^{\pm}(y^{\pm})}{dy^{\pm}}
 \frac{dx^{\pm}(y'^{\pm})}{dy^{\pm}}\partial_{\pm} f(x^{\pm})
\partial_{\pm}
f(x'^{\pm})+\frac{1}{4\pi}\frac{1}{(y^{\pm}-y'^{\pm})^2} \ \ \ \ \
\ \ \
 \eea
and the result is \be \label{trl} :T_{\pm\pm}(y^{\pm}): = \left(
\frac{dx^{\pm}}{dy^{\pm}}\right)^2 :T_{\pm\pm}(x^{\pm}): -
\frac{1}{24\pi}\{ x^{\pm}, y^{\pm} \} \ ,\ee where \be \{ x^{\pm},
y^{\pm}\}= \frac{d^3 x^{\pm}}{dy^{\pm 3}}/ \frac{d x^{\pm }}{d
y^{\pm}} -\frac{3}{2} \left(\frac{d^2 x^{\pm}}{d y^{\pm
2}}/\frac{d x^{\pm }}{d y^{\pm}}\right)^2 \ \ee is the Schwarzian
derivative. Note that the normal ordered operator
$:T_{\pm\pm}(x^{\pm}):$ does not transform  as a tensor. Normal
ordering breaks the classical covariant transformation law under
conformal transformations \be T_{\pm\pm}(y^{\pm}) = \left(
\frac{dx^{\pm}}{dy^{\pm}}\right)^2 T_{\pm\pm}(x^{\pm}) \ . \ee
Indeed, normal ordering requires a selection of modes, and
therefore of coordinates. For instance, $:T_{\pm\pm}(x^{\pm}):$
can be defined from the plane-wave modes $u_w=(4\pi
w)^{-1/2}e^{-iwx^{\pm}}$ and $:T_{\pm\pm}(y^{\pm}):$, instead,
from the modes ${\tilde{u}}_w =(4\pi w)^{-1/2}e^{-iwy^{\pm}}$.

The two-point correlation function $\langle
0_x|:\partial_{\pm}f(y^{\pm})\partial_{\pm}f(y'^{\pm}):|0_x\rangle$
also serves to construct the particle number operator. We start
from the explicit form of it in terms of  creation and
annihilation operators (for simplicity we shall consider only the
right mover sector)
\begin{eqnarray} \label{expno}
 \langle 0_x|:\partial_{y^-_1}f(y^-_1)\partial_{y^-_2}f(y^-_2):|0_x\rangle
&=& \sum_{ji} \left\{ \langle 0_x |
{\stackrel{\rightarrow}{b}}_j^\dagger
{\stackrel{\rightarrow}{b}}_i |0_x \rangle
\left(\partial_{y^-_1}\tilde{u}_i\partial_{y^-_2}\tilde{u}^*_j+
      \partial_{y^-_1}\tilde{u}^*_j\partial_{y^-_2}\tilde{u}_i\right)+\right.\nonumber\\
&+&\left.\left(\langle 0_x |
{\stackrel{\rightarrow}{b}}_j{\stackrel{\rightarrow}{b}}_i |0_x
\rangle \partial_{y^-_1}\tilde{u}_j\partial_{y^-_2}\tilde{u}_i
 +c.c.\right)\right\}
\ . \end{eqnarray}
 Now, instead of taking the
limit $y^{-}_1 \to y^{-}_2$, as in the construction of the stress
tensor, we shall perform the following  transform
\begin{equation}
\int_{-\infty}^{+\infty} dy^-_1dy^-_2
\tilde{u}_k(y^-_1)\tilde{u}^*_{k'}(y^-_2)\langle 0_x
|:\partial_{y^-_1} f(y^{-}_1)\partial_{y^-_2} f(y^{-}_2): |0_x
\rangle \ .
\end{equation}
 We can evaluate this expression in terms of the
particle number operator. To this end we shall use (\ref{expno})
together with the relations
\begin{eqnarray}
\begin{array}{c}
(\tilde{u}_i,\tilde{u}_j)=
-2i\int_{-\infty}^{\infty}dy^-\tilde{u}_i\partial_{y^-}\tilde{u}^{*}_j
=\delta_{ij} \\
(\tilde{u}^{*}_i,\tilde{u}^{*}_j)=
-2i\int_{-\infty}^{\infty}dy^-\tilde{u}^{*}_i\partial_{y^-}\tilde{u}_j
= -\delta_{ij}\\
(\tilde{u}_i,\tilde{u}^{*}_j)=
-2i\int_{-\infty}^{\infty}dy^-\tilde{u}_i\partial_{y^-}\tilde{u}_j
=0
\end{array}
\end{eqnarray}
 The result is as follows \be \int_{-\infty}^{+\infty}
dy^-_1dy^-_2 \tilde{u}_k(y^-_1)\tilde{u}^*_{k'}(y^-_2)\langle 0_x
|:\partial_{y^-_1} f(y^{-}_1)\partial_{y^-_2} f(y^{-}_2): |0_x
\rangle = \frac{1}{4} \langle 0_x |
{\stackrel{\rightarrow}{b}}^\dagger_k
{\stackrel{\rightarrow}{b}}_{k'} |0_x \rangle \ . \ee We then
immediately get an expression for the expectation value of the
particle number operator
${\stackrel{\rightarrow}{N}}_{k}={\stackrel{\rightarrow}{b}}^\dagger_k{\stackrel{\rightarrow}{b}}_k$
associated to the right-moving mode $k$ : \be
 \langle 0_x | {\stackrel{\rightarrow}{N}}_k |0_x \rangle = 4 \int_{-\infty}^{+\infty} dy^-_1dy^-_2
 \tilde{u}_k(y^-_1)\tilde{u}^*_{k}(y^-_2)\langle 0_x
|:\partial_{y^-_1} f(y^{-}_1)\partial_{y^-_2} f(y^{-}_2): |0_x
\rangle . \ee Taking into account (\ref{normalorder}) we obtain
\bea \label{partnumbcorr} \langle 0_x |
{\stackrel{\rightarrow}{N}}_k |0_x \rangle &=& -\frac{1}{\pi}
\int_{-\infty}^{+\infty} dy^-_1
dy^-_2\tilde{u}_k(y^-_1)\tilde{u}^*_{k}(y^-_2)\times  \\ & &
\left[
 \left(\frac{dx^{-}(y^{-}_1)}{dy^{-}}\right)
 \left(\frac{dx^{-}(y^{-}_2)}{dy^{-}}\right)
 \frac{1}{(x^{-}(y^-_1)-x^{-}(y^-_2))^2}-
 \frac{1}{(y^{-}_1-y^{-}_2)^2}\right]  \nonumber \ .\eea
 This expression has a nice physical interpretation. The production
 of quanta, as measured by an observer with coordinates $y^{\pm}$,
 is clearly
 associated to the deviation of the correlations $\langle 0_x
|\partial_{y^-_1} f(y^{-}_1)\partial_{y^-_2} f(y^{-}_2) |0_x
\rangle$ from their corresponding value in the vacuum $|0_y
\rangle$. Moreover, the correlations contributing to the
production of quanta in the mode $k$ are those supported at the
set of points $y^-_1$ and $y^-_2$ where the mode is located. This
is more clear when one introduce finite-normalization wave packet
modes, instead of the usual plane wave modes \be
\tilde{u}_w=\frac{e^{-iwy^-}}{\sqrt{4\pi w}},\ee for which
$(\tilde{u}_w,\tilde{u}_{w'})=\delta(w-w')$\footnote{Note that the
integral $\int_0^{\infty}dw w\langle 0_x |
{\stackrel{\rightarrow}{N}}_w |0_x \rangle$ gives the integrated
flux $\int dy^- \langle 0_x| : T_{--}(y^-): |0_x\rangle $.}. The
wave packet modes can be defined as follows \cite{hawking}
\be\tilde{u}_{jn}=\frac{1}{\sqrt{\epsilon}}\int_{j\epsilon}^{(j+1)\epsilon}
dw e^{2\pi inw/\epsilon}\tilde{u}_w \ , \ee with integers $j \geq
0$, $n$. These wave packets are peaked about $y^-=2\pi
n/{\epsilon}$ with width $2\pi /\epsilon $. Taking $\epsilon$
small ensures that the modes are narrowly centered around $w
\simeq w_{j}=j\epsilon$. Therefore the main contribution to
$\langle 0_x|{\stackrel{\rightarrow}{N}}_{jn}|0_x\rangle $ comes
from correlations, of range similar to the support of the wave
packet, around the point $y^-=2\pi n/{\epsilon}$.

It is interesting to  remark that the difference of two-point
functions in (\ref{partnumbcorr}) at $y_1^-=y_2^-$ is not
singular. In fact, for $y_1^-=y_2^-+\epsilon$ and $|\epsilon|<<1$,
it is proportional to \be
 - 4\pi \langle 0_x| : T_{--}(y^-):
|0_x\rangle -2\pi \frac{d}{dy^-}\langle 0_x| : T_{--}(y^-):
|0_x\rangle  \epsilon +O(\epsilon^2)\ ,\ee which clearly shows a
smooth behaviour at the coincidence limit.

Finally we  mention that the expression (\ref{partnumbcorr}) must
be equivalent, by construction, to that given in terms of
Bogolubov coefficients (\ref{pnbog}). The aim of next sections is
to show that (\ref{partnumbcorr}) offers an interesting
perspective to  understand better the phenomena of particle
production.

\section{Thermal radiation}

As an illustrative example we shall show how the expression
(\ref{partnumbcorr}) reproduces the thermal properties associated
to the conformal transformation \be
\label{rindlertr}x^{\pm}={\pm}{\kappa}^{-1}e^{\pm \kappa y^{\pm}}
. \ee We can think of this transformation as relating the
Minkowskian $x^{\pm}$ and Rindler $y^{\pm}$ null coordinates,
where $\kappa$ is the acceleration parameter (the same relation
holds for the Schwarzschild black hole between the Kruskal and
Eddington-Finkelstein null coordinates with $\kappa=1/4M$). As an
intermediate step we shall first make use of plane waves and work
out an expression for $\langle 0_x |
{\stackrel{\rightarrow}{b}}_w^\dagger
{\stackrel{\rightarrow}{b}}_{w'} |0_x \rangle$ \bea \langle 0_x |
{\stackrel{\rightarrow}{b}}_w^\dagger
{\stackrel{\rightarrow}{b}}_{w'} |0_x \rangle &=&
-\frac{1}{4\pi^2\sqrt{ww'}} \int_{-\infty}^{+\infty} dy^- dy'^-
\left[
 \frac{dx^{-}}{dy^{-}}(y^{-})
 \frac{dx^{-}}{dy^{-}}(y'^{-}) \frac{1}{(x^{-}-x'^{-})^2}\right.
 \nonumber \\ &-&
 \left.\frac{1}{(y^{-}-y'^{-})^2}\right] e^{-iw y^- +iw' y'^-} \ .\eea
Substitution of the relations (\ref{rindlertr}) yields to \be
\langle 0_x | {\stackrel{\rightarrow}{b}}_w^\dagger
{\stackrel{\rightarrow}{b}}_{w'} |0_x \rangle = -\frac{1}{2\pi
w}\delta(w-w')\int_{-\infty}^{+\infty} dz \left[ \frac{\kappa^2
e^{-\kappa z}}{(1-e^{-\kappa z})^2} -\frac{1}{z^2} \right]
e^{-iwz}\ ,\ee where $z=y^- - y'^-$. Evaluation of the integral
gives \be \langle 0_x | {\stackrel{\rightarrow}{b}}_w^\dagger
{\stackrel{\rightarrow}{b}}_{w'} |0_x \rangle= \delta (w-w')
\frac{1}{e^{\frac{2\pi w}{\kappa}}-1}\  . \ee The delta function
leads to a divergent result for the emitted number of particles
$\langle 0_x | {\stackrel{\rightarrow}{N}}_w |0_x \rangle =
\langle 0_x | {\stackrel{\rightarrow}{b}}_w^\dagger
{\stackrel{\rightarrow}{b}}_{w} |0_x \rangle$. As usual, this
divergence can be cured  introducing a basis of
finite-normalization wave packet modes. If we evaluate, instead,
$\langle 0_x | {\stackrel{\rightarrow}{N}}_{jn} |0_x \rangle$,
using again (\ref{partnumbcorr}), it turns out that \bea \langle
0_x | {\stackrel{\rightarrow}{N}}_{jn} |0_x \rangle &=&
\frac{1}{\epsilon} \int_{j\epsilon}^{(j+1)\epsilon}dw e^{2\pi iw
n/\epsilon}\int_{j\epsilon}^{(j+1)\epsilon}dw' e^{-2\pi
iw'n/\epsilon} \langle 0_x | {\stackrel{\rightarrow}{b}}_w^\dagger
{\stackrel{\rightarrow}{b}}_{w'} |0_x \rangle  \nonumber \\
&=& \frac{1}{\epsilon}
\int_{j\epsilon}^{(j+1)\epsilon}dw\frac{1}{e^{8\pi Mw} - 1}=
\frac{1}{e^{8\pi M w_{j}} - 1}\ , \eea where in the last step we
have assumed that the wave packets are sharply peaked around the
frequencies $w_{j}$. This  corresponds to the Planckian spectrum
of radiation at the temperature $T=\frac{\kappa}{2\pi}$. Similar
results hold for the left mover sector. Evaluation of the
expectation value of the stress tensor using (\ref{trl}), taking
into account that $ \langle 0_x| :T_{\pm\pm}(x^{\pm}): |0_x\rangle
=0$, gives \be \langle 0_x| :T_{\pm\pm}(y^{\pm}): |0_x\rangle =
\frac{ \kappa^2}{48\pi}= \frac{ \pi T^2}{12}\ .\ee This is nothing
else but the stress tensor corresponding to a two dimensional
thermal bath of radiation at the temperature $T$\footnote{We must
remark, nevertheless, that the covariant  quantum stress tensor
\cite{birrelldavies} $ \langle 0_x| T_{\pm\pm}(y^{\pm})
|0_x\rangle \equiv \langle 0_x| :T_{\pm\pm}(y^{\pm}): |0_x\rangle
-(12\pi )^{-1}(\partial_{\pm}\rho \partial_{\pm}\rho-
\partial_{\pm}^2\rho)$, where the metric is $ds^2=-e^{2\rho}dy^+dy^-=-e^{\kappa(y^+-y^-)}dy^+dy^-$, vanishes.
Despite  the existence of particle production in Rindler space
($\langle 0_x | {\stackrel{\rightarrow}{N}}_{jn} |0_x \rangle \neq
0 \neq  \langle 0_x| :T_{\pm\pm}(y^{\pm}): |0_x\rangle$), the
vanishing of the expectation values of the covariant stress tensor
operator $\langle 0_x| T_{\pm\pm}(y^{\pm}) |0_x\rangle=0$ implies
the absence of backreaction effects on the background flat
metric.}.

Note that our derivation of the Planckian spectrum bypasses the
explicit evaluation of the Bogolubov coefficients. Instead, it is
based on the explicit form of the two-point correlation function,
and the evaluation of the corresponding integral leads directly to
the thermal result.
\\

\section{Mobius transformations}

We shall now analyse the case associated to the Mobius
transformations \be \label{mobius} x^{\pm} \to y^{\pm}=
\frac{a^{\pm}x^{\pm}+b^{\pm}}{c^{\pm}x^{\pm}+d^{\pm}} \ee where
$a^{\pm}d^{\pm}-b^{\pm}c^{\pm}=1$. These form the so-called global
conformal group ($(SL(2,R) \otimes SL(2,R))/Z_{2} \approx
SO(2,2)$) . The physical meaning of these transformations can be
found in \cite{ginsparg}. In addition, a nice physical
interpretation of the special conformal transformations was given
in terms of a uniformly accelerating mirror \cite{daviesfulling,
birrelldavies}.

The Mobius transformations  have the property of giving a
vanishing Schwarzian derivative. Therefore, under the action of
the Mobius transformations the flux of radiation in the vacuum
$|0_x\rangle$ for the observer $\{y^{\pm}\}$ vanishes \be
\label{noflux} \langle 0_x|:T_{\pm\pm}(y^{\pm}):|0_x \rangle =0\
.\ee
 Moreover, since the two-point function
(\ref{correlatordfdf}) is invariant under (\ref{mobius}) it is
clear from (\ref{partnumbcorr}) that the expectation value of the
particle number operator also vanishes \be \label{nopart} \langle
0_x | {\stackrel{\rightarrow}{N}}_k |0_x \rangle =0=\langle 0_x |
{\stackrel{\leftarrow}{N}}_k |0_x \rangle \ ,\ee irrespective of
the particular mode basis. This is indeed what we expect in the
context of CFT, since the vacuum is invariant under Mobius
transformations (see \cite {aldaya} for a different approach).
However, this conclusion is not obvious from the point of view of
Bogolubov coefficients. Let us consider those Mobius
transformations which are not dilatations nor Poincar\'e such as
\be \label{motra} x^-= -\frac{1}{a^2 y^-}\ , \ee where $a$ is an
arbitrary constant. We mention that this transformation originally
appeared in the moving-mirror model of Davies and Fulling
\cite{daviesfulling} (the parameter $a^2$ is related to the
acceleration of the mirror) and more recently in the analysis of
extremal black holes \cite{liberati, gao}, where it gives the
(leading order) relation between the Eddington-Finkelstein and
Kruskal coordinates (which is instead given by (\ref{rindlertr})
in the case of Schwarzschild and non-extremal
Reissner-nordstr\"om),
  and in the late-time behaviour of evaporating near-extremal
Reissner-Nordstrom black holes \cite{fnno}.
The Bogolubov coefficients associated to the standard plane wave
basis are
  \bea \label{albemo}
\alpha_{ww'} &=& \frac{1}{2\pi}\sqrt{\frac{w}{w'}}  \int
_{-\infty}^{+\infty} dy^{-}
 e^{-iwy^- -iw'/a^2 y^- } \nonumber \\   \beta_{ww'}&=& -
\frac{1}{2\pi}\sqrt{\frac{w}{w'}}  \int_{-\infty}^{+\infty} dy^{-}
 e^{-iwy^- +iw'/a^2 y^- } \ . \eea These integrals do not
converge, as it usually happens for the plane wave basis.
Therefore one should  introduce  wave packets. The Bogolubov
coefficients can be then computed from the expressions \bea
\begin{array}{c} \beta_{jn,w'}=-(\tilde{u}_{jn}, u_{w'}^*)=
2i\int_{-\infty}^{+\infty}dy^-\tilde{u}_{jn}\partial_{y^-}u_{w'}\ , \\
\alpha_{jn,w'}=-(\tilde{u}_{jn}, u_{w'})=
-2i\int_{-\infty}^{+\infty}dy^-\tilde{u}_{jn}\partial_{y^-}u_{w'}^*
 \ , \end{array}\ \eea where
\bea  u_{w'}&=& \frac{1}{\sqrt{4\pi w'}}e^{-iw'x^-(y^-)}\nonumber  \\
\tilde{u}_{jn}&=&\frac{1}{\sqrt{\epsilon}}\int_{j\epsilon}^{(j+1)\epsilon}
dw e^{2\pi inw/\epsilon}\tilde{u}_w   \ .  \eea Since $
\tilde{u}_w= \frac{1}{\sqrt{4\pi w}}e^{-iwy^-}$ we have \be
\tilde{u}_{jn}=\frac{1}{\sqrt{4\pi \epsilon
w_{j}}}e^{iw_{j}L}\frac{\sin L\epsilon/2}{L/2} \ , \ee where \be
L= \frac{2\pi n }{\epsilon} -y^- \ . \ee We get  then \bea
\beta_{jn,w'}&=&\frac{1}{\pi \sqrt{\epsilon}}
\sqrt{\frac{w'}{w_j}} \int_{-\infty}^{+\infty}dy^- \frac{\sin
L\epsilon/2}{a^2(y^{-})^2L}e^{iw_{j}L}e^{iw'/a{^2}y^-} \nonumber \\
\alpha_{jn,w'}&=&-\frac{1}{\pi \sqrt{\epsilon}}
\sqrt{\frac{w'}{w_j}} \int_{-\infty}^{+\infty}dy^- \frac{\sin
L\epsilon/2}{a^2(y^{-})^2L}e^{iw_{j}L}e^{-iw'/a{^2}y^-} \ . \eea
According to our previous discussion the first integral above
should vanish, to agree with the result obtained using the Mobius
invariance of the two-point correlation function  \be \langle 0_x
| {\stackrel{\rightarrow}{N}}_{jn} |0_x \rangle =
\int_{0}^{+\infty}dw' |\beta_{jn,w'}|^2 =0 \ . \ee However, to
show that the  first integral vanishes is not easy, due to the
singularity at $y^-=0$.

Summarizing, the absence of particle production is immediate
according to (\ref{partnumbcorr}), but requires a lengthy
elaboration using the Bogolubov coefficients. We regard this as an
indication of the advantage of using the expression
(\ref{partnumbcorr}) to analyze the production of quanta. At this
respect we want to remark that  the analysis leading to the
expression (\ref{partnumbcorr}) is based on the use of correlation
functions of the ``primary" field $\partial_{\pm} f$, rather than
$f$ itself. This avoids the infrared divergence of the scalar
field in two dimensions. In fact, \be \langle 0_x |f(x)f(x')|0_x
\rangle = -\frac{\hbar}{4\pi} \left ( 2\gamma + \ln \lambda
^2(x-x')^2 \right )\ , \ee where $\gamma$ is the Euler constant
and $\lambda$ is an infrared cut-off for frequencies. This
infrared difficulty is cured when one considers instead
correlations of the field $\partial_{\pm}f$ (see
(\ref{correlatordfdf})). In contrast, the Bogolubov coefficients
are defined using mode solutions of the field $f$ itself.
Therefore, it should not be a complete surprise that the result
(\ref{nopart}), which is straightforward using
(\ref{partnumbcorr}), is not so obvious in terms of the Bogolubov
coefficients.

\section{Interpretation in terms of moving mirrors}

All the above discussion can be reinterpreted in terms of the
so-called moving-mirror analogy. The idea is the following.
Instead of having a two-dimensional flat spacetime with two
different sets of modes $(u_{i}(x^-),v_{i}(x^+))$ and
$(\tilde{u}_{i}(y^-), \tilde{v}_{i}(y^+))$ where the coordinates
$y^{\pm}$ and $x^{\pm}$ are related by a conformal transformation:
\bea
y^- &=& y^-(x^-), \nonumber \\
y^+ &=& y^+(x^+) \ , \eea one can introduce a boundary in the
spacetime to produce the same physical consequences. The effect of
the boundary is to disturb the modes in such a way that modes that
at past null infinity behave as $(u_{i}(x^-),v_{i}(x^+))$, once
evolved to future null infinity will take a form similar to
$(\tilde{u}_{i}(y^-), \tilde{v}_{i}(y^+))$. This is the main
property of a mirror model \cite{birrelldavies, carlitz-willey,
parentani}: it can nicely mimic the physics in a non-trivial
background (i.e. Hawking radiation in a black hole geometry), or
the effect of having two different physically relevant sets of
modes in a fixed background (as in the Fulling-Unruh
construction).

The basic ingredient to define a moving mirror model is the
introduction of a (time-dependent) reflecting boundary in the
space such that the field is assumed to satisfy the boundary
condition $f=0$ along its worldline. It is convenient to
parametrize the trajectory of the mirror in terms of null
coordinates
\begin{equation}
x^+=p(x^-) \ . \end{equation} Therefore the boundary condition is
just
\begin{equation}
f(x^-, x^+=p(x^-))=0.
\end{equation}
A null ray at fixed $x^+$ which reflects off the mirror becomes
a null ray of fixed $x^-$. The concrete relation between the
coordinates of this null ray is given by the mirror's trajectory
$x^+=p(x^-)$. In terms of mode functions it is easy to construct
plane wave solutions of the equation $\nabla^2 f =0$ vanishing on
the worldline of the wall:
\begin{equation}
u^{in}_w=\frac{1}{\sqrt{4\pi w}}(e^{-iwx^+}-e^{-iwp(x^-)}) \ .
\end{equation}
They represent a positive frequency wave $e^{-iwx^+}$, coming from
$I^-_R$, that reflects on the curve $x^+=p(x^-)$ and becomes an
outgoing wave $e^{-iwp(x^-)}$, which in general is not a pure
positive frequency wave at $I^+_R$, but rather a superposition of
positive and negative frequency components. In addition we have
also modes representing a pure outgoing positive frequency wave
$e^{-iwx^-}$ at $I^+_R$ which is produced by the reflection of a
wave $e^{-ip^{-1}(x^+)}$ from $I^-_R$
\begin{equation}
u^{out}_w=\frac{1}{\sqrt{4\pi w}}(e^{-iwx^-}-e^{-iwp^{-1}(x^+)}) \
.
\end{equation}
The above two sets of modes are the natural mode basis for
inertial observers at $I^-_R$ and $I^+_R$ and allow to define the
corresponding IN and OUT vacuum states. This concerns the dynamics
of the field at the right hand side of the mirror. Similar basis
can be constructed to describe the dynamics at the left of the
mirror, but we shall restrict, as usual, to the right region.
Moreover, we can construct wave packet basis from the plane wave
modes and re-derive the same results obtained in section 3. The
expectation value of the particle number operator in the mode $k$
is given by \bea \label{Nmirrors}
 \langle 0_{in} |N^{out}_k |0_{in} \rangle &=& 4 \int_{I^+_R} dx^-_1dx^-_2
 u^{out}_k(x^-_1)u^{out*}_{k}(x^-_2)\langle 0_{in}
|:\partial_{x^-_1} f(x^{-}_1)\partial_{x^-_2} f(x^{-}_2): |0_{in}
\rangle \nonumber  \\
&=& -\frac{1}{\pi} \int_{I^+_R} dx^-_1 dx^-_2 u^{out}_k(x^-_1)
u^{out*}_{k}(x^-_2)\times \\ & & \left[
 \frac{p'(x^-_1)p'(x^-_2)}{(p(x^-_1)-p(x^-_2))^2}-
 \frac{1}{(x^{-}_1-x^{-}_2)^2}\right] \nonumber  \ , \eea
and the flux of energy radiated to the right is  given by the
Schwarzian derivative \be \label{mirrorflux} \langle
0_{in}|:T_{--}(x^-):|0_{in}\rangle = - \frac{1}{24\pi}\{ p(x^-) ,
x^{-} \} \ .\ee The results concerning thermal radiation obtained
in section 4 can be rederived in this context by considering the
mirror trajectory $x^+=-\kappa^{-1}e^{-\kappa x^-}$.

We shall now illustrate our previous discussion on Mobius
transformations with the use of the moving-mirror analogy.

\begin{center}
TWO HYPERBOLIC MIRRORS
\end{center}

Our first example will be a mirror with two hyperbolic branches
\begin{equation}
p(x^-)=-\frac{1}{a^2x^-} \ . \end{equation} One branch with
$x^-<0$ and the other with $x^->0$ (see Fig.1.)

\begin{center}
\begin{tabular}{c}
\includegraphics[width=0.45\textwidth,angle=-90]{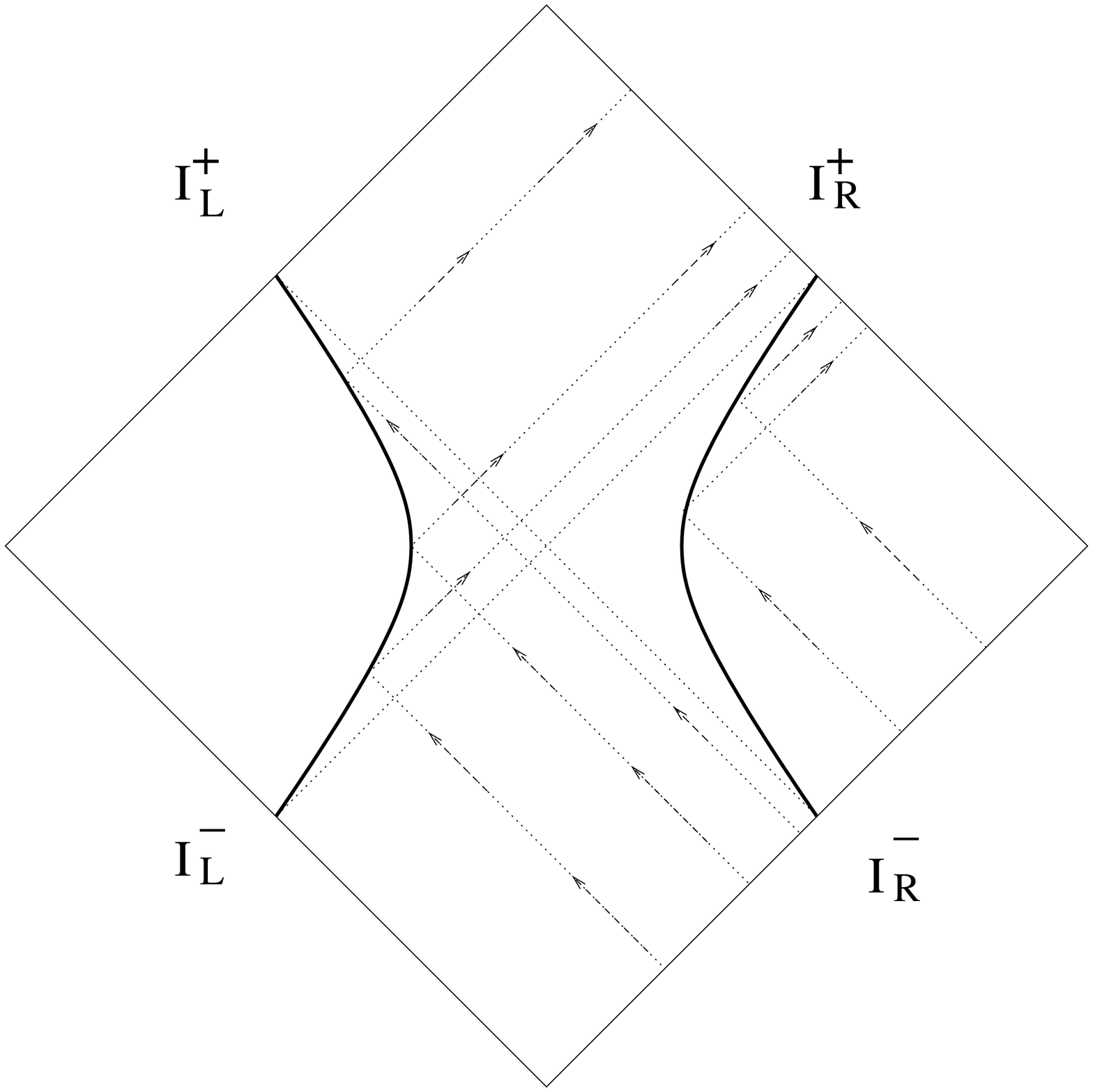}\\
{\scriptsize \textbf{Fig.1} A mirror with two hyperbolic
branches.}
\end{tabular}
\end{center}

The above function $p(x^-)$ can be regarded as associated to a
special conformal transformation of coordinates. Note that all the
modes supported on $I^-_R$ are reflected to $I^+_R$. The IN modes
supported on the interval $ x^+ \in ]-\infty,0[$ reach $I^+_R$ on
$x^- \in ]0,\infty[$; the IN modes supported on the interval $x^+
\in ]0,\infty[$ reach $I^+_R$ on  $x^-\in ]-\infty,0[$. In this
way, the correlations existing between positive and negative $x^+$
are transferred to correlations between positive and negative
$x^-$. Moreover, since the two-point correlation function on
$I^+_R$ is the same as that of the vacuum
\begin{eqnarray}
\langle
0_{in}|\partial_{x^-}\phi(x^-_1)\partial_{x^-}\phi(x^-_2)|0_{in}\rangle
&=&-\frac{1}{4\pi}\frac{p'(x^-_1)p'(x^-_2)}{(p(x^-_1)-p(x^-_2))^2}\nonumber \\
 &=& -\frac{1}{4\pi}\frac{1}{(x^-_1-x^-_2)^2} \ \makebox{   for all
} x^-_1,x^-_2 \ , \end{eqnarray} there is neither particle
production $\langle 0_{in} |N^{out}_k |0_{in} \rangle =0$ nor
energy flux $\langle 0_{in}|:T_{--}(x^-):|0_{in}\rangle =0$.

\begin{center}
 HYPERBOLIC MIRROR ACCELERATING FROM REST
\end{center}

Let us consider a mirror that from rest accelerates to the left
following an hyperbolic trajectory, i.e.,
\begin{eqnarray}
p(x^-)=\left\{\begin{array}{cc}
                 x^- & \makebox{ if } x^-\le 0 \\
                 \frac{x^-}{1+a^2x^-} & \makebox{ if } x^-\ge 0 .
                 \end{array}\right.
\end{eqnarray}
We can write this trajectory in the following compact notation
\begin{equation}\label{trajectori1b}
p(x^-)= x^-\theta(-x^-) + \frac{x^-}{1+a^2x^-}\theta(x^-).
\end{equation}

\begin{center}
\begin{tabular}{c}
\includegraphics[width=0.45\textwidth,angle=-90]{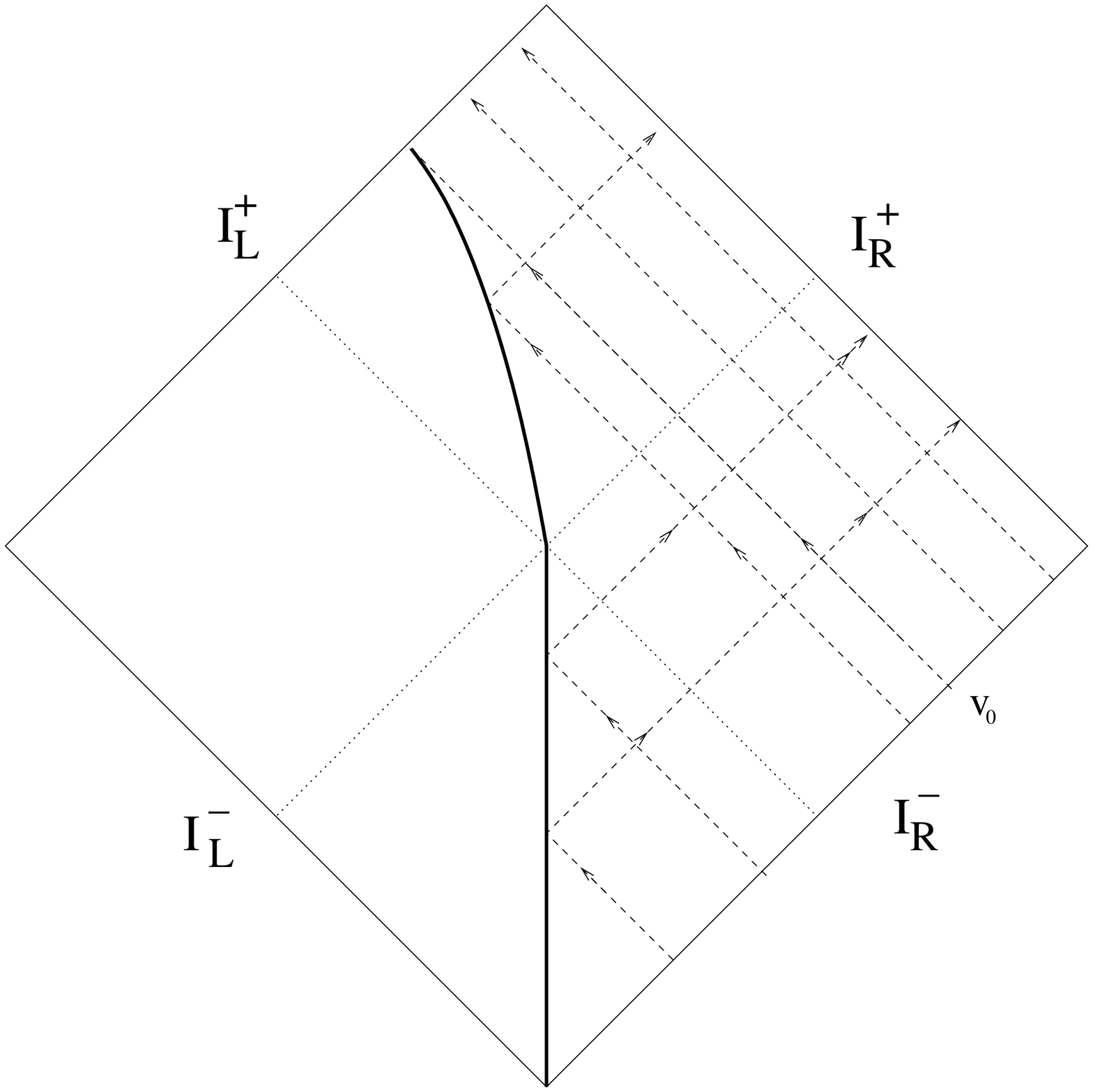}\\
{\scriptsize \textbf{Fig.2} Hyperbolic mirror accelerating from
rest.}
\end{tabular}
\end{center}

In this case (see Fig. $2$) the mirror starts at $i^-$ and ends up
on $I^+_L$. This gives rise to the appearance of a horizon: the IN
modes in the range $x^+\in [1/a^2,\infty[$ do not reach $I^+_R$.
On the other hand the modes supported in the range $x^+\in
[-\infty ,1/a^2[$, upon reflection off the mirror, will reach
$I^+_R$. The corresponding correlation function along $I^+_R$
becomes
\begin{eqnarray}\label{eq:2p-hyp}
\left<0_{in}|\partial_{x^-}\phi(x_1)\partial_{x^-}\phi(x_2)|0_{in}\right>&=&\left\{\begin{array}{cc}
-\frac{1}{4\pi}\frac{1}{(x^-_1-x^-_2)^2} & \makebox{ if } x^-_1
\times x^-_2>0
\\
. \\
-\frac{1}{4\pi}\frac{1}{(x^-_1-x^-_2+a^2x^-_1x^-_2)^2} & \makebox{
if } x^-_1 \times x^-_2<0 \ . \end{array}\right.
\end{eqnarray}
From this we learn that the identity transformation for $x^-<0$
and the ``single-branch'' special conformal transformation for
$x^->0$ make it impossible to distinguish the IN vacuum  state
from the OUT vacuum by means of measurements restricted to
$x^-_1,x^-_2>0$ or to $x^-_1,x^-_2<0$. Only the mixed correlations
$x^-_1>0$, $x^-_2<0$ and $x^-_1<0$, $x^-_2>0$ allow to distinguish
the IN vacuum from the OUT vacuum. Moreover, even though the flux
is zero for $x^-<0$ and $x^->0$ (since in these regions the
normal-ordered two-point correlation function is identically
zero), there is a divergence at $x^-=0$. The evaluation of the
Schwarzian derivative for the trajectory (\ref{trajectori1b})
gives
\begin{equation}
\langle 0_{in}|T_{--}|0_{in}\rangle = \frac{a^2}{12\pi}\delta(x^-)
\ .
\end{equation}
The origin of this non-vanishing flux at $x^-=0$ can be attributed
to the deviation of the correlation function (for $x^-_1<0$ and
$x^-_2>0$) from that of the vacuum. This is not surprising, since
the mirror has undergone a sudden acceleration just at $x^-=0$.
The same reason underlies the (non-vanishing) production of
quanta, which according to eq. (\ref{Nmirrors}) turns out to be:
\begin{eqnarray}
<0_{in}|N_k|0_{in}>&=&-\frac{1}{\pi}\int_{-\infty}^0 dx^-_1
\int_0^{\infty}dx^-_2 (u^{out}_k(x^-_1){u^{out}}^*_k(x^-_2)+
u^{out}_k(x^-_2){u^{out}}^*_k(x^-_1))\times \nonumber
\\
&&\left[\frac{1}{(x^-_1-x^-_2+a^2x^-_1x^-_2)^2}-\frac{1}{(x^-_1-x^-_2)^2}\right].
\end{eqnarray}

We observe that for modes $k$ supported in the region $x^->0$ the
expectation value $<0_{in}|N_k|0_{in}>$ vanishes. Therefore, a
particle detector which is switched on at late times $x^->>0$ ( or
early times $x^-<<0$) , will never detect the emission of quanta,
since there $\langle 0_{in}|N_k|0_{in}\rangle =0$. The detection
of quanta will only take place through the region $x^-=0$ where
the flux is non-vanishing. In other words, the measured quanta
needs to correspond to a wave packet mode with support around the
point $x^-=0$.

\begin{center}
 SINGLE-BRANCH HYPERBOLIC MIRROR
\end{center}

Let us now consider again, as in the first example, a pure
hyperbolic mirror, but this time only a single branch
\begin{equation}\label{trajectoripure1b}
p(x^-)= - \frac{1}{a^2x^-}\theta(x^-) \ . \end{equation} For
$x^-<0$ there is no reflecting wall. This case is more involved
since Right and Left are not disconnected. According to Fig.$3$,
the IN modes reaching $I^+_R$ are of two types: those coming from
the $(-\infty,0)$ segment of $I^-_R$ and those coming from the
$(-\infty,0)$ segment of $I^-_L$.
\begin{center}
\begin{tabular}{c}
\includegraphics[width=0.45\textwidth,angle=-90]{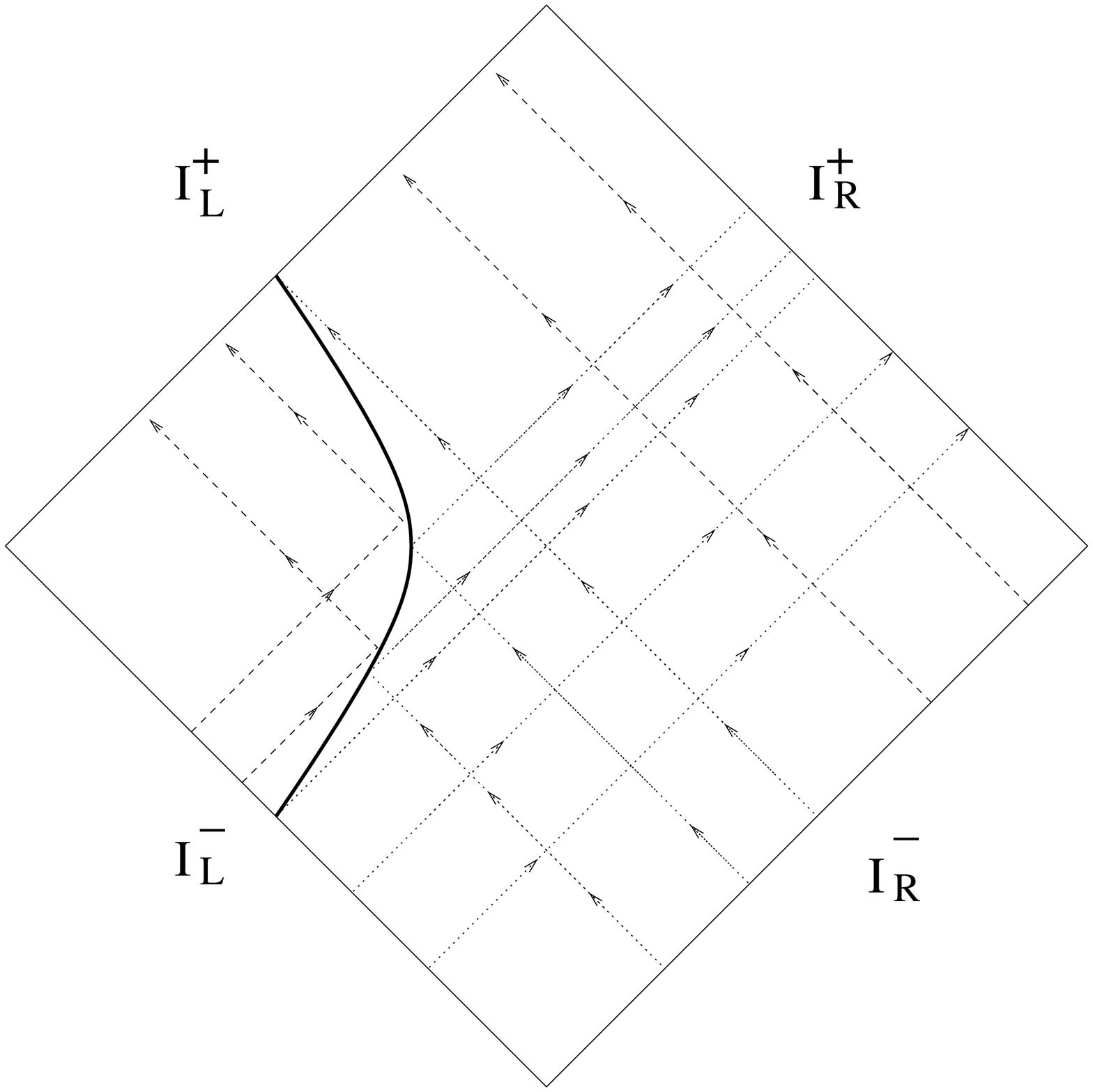}\\
{\scriptsize \textbf{Fig.3} A single-branch hyperbolic mirror.}
\end{tabular}
\end{center}

 Since there are no correlations between $I^-_R$ and
$I^-_L$, it is easy to see that the two-point correlation function
on $I^+_R$ is given by
\begin{eqnarray}\label{eq:2p-halfMob-u}
\left<0_{in}|\partial_{x^-}\phi(x^-_1)\partial_{x^-}\phi(x^-_2)|0_{in}\right>&=&\left\{\begin{array}{cc}
-\frac{1}{4\pi}\frac{1}{(x^-_1-x^-_2)^2} & \makebox{ if }
x^-_1,x^-_2>0 \makebox{ or } x^-_1,x^-_2<0\\ \\0 & \makebox{ if }
x^-_1<0, x^-_2>0 \makebox{ or } x^-_1>0, x^-_2<0 . \ \ \ \ \ \ \ \
\end{array}\right.
\end{eqnarray}
 The particle production is then
given by the expression
\begin{eqnarray}\label{N1branch}
<0_{in}|N_k|0_{in}>&=&-\frac{1}{\pi}\int_{-\infty}^0 dx^-_1
\int_0^{\infty}dx^-_2 (u^{out}_k(x^-_1){u^{out}}^*_k(x^-_2)+
u^{out}_k(x^-_2){u^{out}}^*_k(x^-_1))\times \nonumber
\\
&&\left[-\frac{1}{(x^-_1-x^-_2)^2}\right] \ . \end{eqnarray} We
observe that the non-vanishing contribution comes from the OUT
vacuum correlations between positive and negative points, since
they cannot be canceled out by the correlations of the IN vacuum
state, as it happens instead for the pair of points $x^-_1,x^-_2
>0$, or $x^-_1,x^-_2 <0$. Due to this there is a divergent flux of
energy concentrated at the point $x^-=0$, dividing the two
uncorrelated regions with respect to the IN vacuum state. This
divergence is then even more drastic than that found in the
previous example, for which the correlations between points
$x^-_1>0$ and $x^-_2<0$ are diminished with respect to the OUT
vacuum, but are non-zero. We can also infer the same type of
conclusions for the production of quanta. At late, or early times,
we will never detect quanta, since the required modes are those
with support covering the point $x^-=0$ where the energy flux is
concentrated. This result agrees with that obtained in terms of
particle detectors \cite{grove}.

Let us now compare this analysis with the interpretation of
\cite{daviesfulling, birrelldavies} carried out employing naively
the Bogolubov coefficients. As we have already remarked the
Bogolubov coefficients associated to special conformal
transformations for plane waves involve ill-defined integrals.
Restriction of the special conformal transformation to one single
branch still produces ill-defined expressions for the Bogolubov
coefficients: \bea \label{albemo} \alpha_{ww'} &=&
\frac{1}{2\pi}\sqrt{\frac{w}{w'}} \int _{0}^{+\infty} dy^{-}
 e^{-iwy^- -iw'/a^2 y^- }, \nonumber \\   \beta_{ww'}&=& -
\frac{1}{2\pi}\sqrt{\frac{w}{w'}}  \int_{0}^{+\infty} dy^{-}
 e^{-iwy^- +iw'/a^2 y^- } \ . \eea
 Following the original work on the moving-mirror system \cite{daviesfulling},
 one can evaluate these integrals using a
 Wick rotation (i.e., integrating along the imaginary axis)
 \footnote{This has been recently criticized in \cite{gao}.}.
The results are \cite{daviesfulling, birrelldavies, parentani}
\bea \label{albemo2} \alpha_{ww'} &=&
\frac{1}{a\pi} K_1 (2i\sqrt{ww'/a^2})\ , \nonumber \\
\beta_{ww'}&=&\frac{i}{a\pi} K_1(2\sqrt{ww'/a^2})\ , \eea where
$K_1$ is a modified Bessel function\footnote{Notice that if the
integral is from $-\infty$ to $+\infty$ (i.e., the two branches of
the special conformal transformation) there is cancelation between
both branches and the final result is $\beta_{ww'}=0$.}. The
nonvanishing of the $\beta_{ww'}$ coefficients was interpreted
\cite{daviesfulling, birrelldavies} as an indication of the
existence of particle production even in the absence of energy
fluxes. However, due to the term $1/\sqrt{ww'}$ in the asymptotic
form of $K_1$ for small frequencies, the quantity
\begin{equation}
\langle 0_{in}|N_w|0_{in}\rangle=
\int_{0}^{+\infty}dw'|\beta_{ww'}|^2 \  \end{equation} diverges.
As usual, this should be cured introducing wave packet modes, but,
as far as we know, an explicit calculation has never been
performed (see also \cite{gao}). It is therefore difficult to find
a clear physical picture for the time distribution of the emitted
quanta and draw definite conclusions within the approach of
Bogolubov coefficients.

However, we can easily match with our conclusions from the
expression (\ref{N1branch}) if we make the following reasoning.
The IN vacuum at $I^-_{R}$ can be expanded in terms of correlated
Rindler particles between both sides of the line $x^+=0$. For this
we can introduce the unconstrained coordinates $\kappa
y^+=-\ln(-\kappa x^+)$, for $x^+<0$ , and $\kappa z^+=\ln\kappa
x^+$, for $x^+>0$. $\kappa$ is an arbitrary positive constant
which plays only an auxiliary role in the discussion. In a
parallel way we can introduce the coordinates $\kappa
y^-=\ln\kappa x^-$, for $x^->0$ , and $\kappa z^-=-\ln (-\kappa
x^-)$, for $x^-<0$. The Rindler modes $e^{-iwy^+}$ and $e^{-iwy^-}$
are related, due to the reflection, by means of the special
conformal transformation. But in the new coordinates the special
conformal transformation (\ref{trajectoripure1b}) is ``inertial"
\begin{equation}
\kappa y^+ = -\ln\frac{\kappa^2}{ a^2} +\kappa y^- \ .
\end{equation} The corresponding $\beta$ coefficients are then
clearly zero. The restriction of the IN vacuum to the segment
$x^+<0$ is a mixed state. Moreover, in terms of the Rindler
coordinate $y^+$, this mixed state takes the form of a thermal
state (of Rindler particles) with temperature $T=\kappa /2\pi$.
The OUT vacuum state is also mixed when restricted to $x^->0$, and
a thermal state with respect to the Rindler coordinate $y^-$. Due
to the vanishing of the $\beta$ coefficients the thermal
description of the IN vacuum, when restricted to $x^+<0$, is
reflected without distortion by the mirror, producing the same
thermal state at $I^+_{R}$. This thermal state is supported in the
region $x^->0$, and has no correlations with the region $x^-<0$.
As we have already noted, this thermal state is perceived, for the
inertial observer at $I^+_{R}$ using the coordinate $x^-$, the
same as the Minkowski OUT vacuum for measurements restricted to
the region $x^->0$. Therefore, in the region $x^->0$ neither
particle production nor energy flux can be detected.

\section{Conclusions}

In this note we have analyzed the particle production due to
conformal transformations or, equivalently, due to reflections on
moving mirrors, based on a viewpoint different from the standard
approach. We were motivated by the fact that, under special
conformal transformations, the vanishing of the $\beta$
coefficients is not trivial, despite the fact that the invariance
of the vacuum under Mobius transformations is one of the
``postulates" of CFT. The expression for the particle production
that we analyze here (eq.(\ref{partnumbcorr})) immediately
clarifies this aspect. This is so because it emphasizes that the
deviation of the two-point correlation function from that of the
vacuum, weighted by wave packets of a definite mode, is the source
of the production of quanta of the corresponding mode. We have
first shown that the proposed expression for the particle
production works nicely in recovering, in a simple way, the
standard thermal radiation in Rindler space. We have also revised,
from this point of view, the moving-mirror systems with different
examples of hyperbolic trajectories. We have pointed out the close
relation between the production of quanta and energy, which
contrasts early conclusions based on ill-defined expressions for
the Bogolubov coefficients.
\\ \noindent Finally we remark that are we not criticizing the Bogolubov approach
in favor of the approach presented here. Both approaches are
different ways of  measuring the same physical quantity.
\\ \\ \noindent
Note added: After completion of this work we were informed that in \cite{hkp}
particle production is also investigated without
making use of Bogolubov coefficients in a cosmological scenario.

\section*{Acknowledgements}
A.F. and J.N-S thank V. Frolov, S. Gao and A. Zelnikov and G.O.
thanks L. Parker for very useful discussions. This research has
been partially supported by the research grants
BFM2002-04031-C02-01 and BFM2002-03681 from the Ministerio de
Ciencia y Tecnologia (Spain), EU FEDER funds and the INFN-CYCIT
Collaborative Program. We also thank  an anonymous referee for
interesting comments.

\end{document}